\documentclass[amsmath,amssymb,nofootinbib]{revtex4}

\usepackage{latexsym,comment,verbatim}
\usepackage{amssymb,multirow}
\usepackage{amsfonts}
\usepackage{amsmath,color}
\usepackage{graphicx}
\usepackage{bm}

\def\ber{\begin{eqnarray}}
\def\eer{\end{eqnarray}}
\def\beq{\begin{equation}}
\def\eeq{\end{equation}}

\def\ed{\end{document}}

\begin{document}

\title{Physics for the masses: teaching Einsteinian gravity in primary school}

\author{Matteo Luca Ruggiero}
\email{matteo.ruggiero@polito.it}
\affiliation{Politecnico di Torino, Torino (Italy )and INFN, Laboratori Nazionari di Legnaro, Legnaro  (Italy)}

\author{Sara Mattiello}
\affiliation{Universit\`a degli Studi di Torino, Torino (Italy)}

\author{Matteo Leone}
\affiliation{Universit\`a degli Studi di Torino, Torino (Italy)}

\date{\today}

\begin{abstract}
 {Why is modern physics still today, more than 100 years after its birth, the privilege of an elite of scientists} and unknown for the great majority of citizens? The answer is simple, since modern physics is in general not present in the standard physics curricula, except for some general outlines, in the final years of some secondary schools. But,  is it possibile to teach modern physics in primary school? Is it effective? And, also,  {is it engaging for students?}
These are the simple questions which stimulated our research, based on an intervention performed in the last year of Italian primary school,  focused on teaching gravity, according to the Einsteinian approach {in the spirit of the Einstein First project, an international collaboration  which aims to teach school age children the concepts of modern physics. The outcomes of our research study are in agreement with previous findings obtained in Australian schools, thus they contribute to validate them and show that there is no cultural effect, since the approach works in different education systems.  {Finally, our results are relevant} also in terms of retention and prove that  the students involved really understand the key ideas. }
 \end{abstract}

\maketitle

\section{Introduction}\label{sec:intro}

Einstein's theory of gravity, general relativity, is our best model of gravity, since  it passed with flying colors many observational tests \cite{will2014confrontation}, however, together with quantum mechanics, the other pillar of modern physics, it is usually out of school curricula except for their general outlines in the final years of some secondary schools. 
These theories are important not only because of the change they provoked in the paradigm of physics, but also for their applications which have a huge impact on everyday life, but are generally    beyond the grasp of the great majority of citizens. Recent results show that  it is possible and also effective to introduce modern physics early in schools: the Einstein First project \cite{pitts2014exploratory,kaur2017teaching1,kaur2017teaching2,kaur2017teaching3,kaur2017gender,kaur2017evaluation,foppoli2018public,choudhary2018can} aims to develop learning sequences on modern physics as early as the primary school years. According to this approach, the introduction of language and concepts of the modern understanding of space, time, matter and radiation at an early age could prevent conceptual conflicts between Newtonian physics and Euclidean geometry  (that are commonly taught  at school) and the completely different paradigms of modern physics. 

It is important to emphasize that this new approach does not suggest to completely abandon the paradigm of classical physics, because it is important to understand many phenomena; rather, it is suggested that classical physics can be introduced \textit{after} learning the  language and basic concepts of modern physics.
The results of the educational programs inspired by Einstein First project suggest that this approach could be fruitful both at the primary and at the secondary school level. In particular, the preliminary results obtained by these programs suggest that an  early exposure to the modern paradigms could be effective in increasing pupils' attitudes towards science  (see eg. \cite{pitts2014exploratory, kaur2017teaching3, choudhary2018can}) and in promoting the learning of modern physics  \cite{pitts2014exploratory, foppoli2018public}.

In this paper, with the goal of exploring the educational power of a suitably designed modern physics approach at the primary school level,
we report the results of a research study performed  on a sample of 10 to 11 yrs old Italian primary school students in which we introduced some concepts of the theory of relativity, such as the role of the reference frame, the velocity addition law, the peculiar characteristics of light propagation, the role of simultaneity and, eventually, gravitational interactions according to the Einsteinian picture. In particular, here we will focus  on teaching Einstenian gravity, while the rest of the results will be addressed in a subsequent analysis.  Actually, the first research in this field dates back to the work of \citet{haddad1972relationship}, which reported evidences that pupils in primary school can understand concepts of relativity, at least at the knowledge level, according to Bloom taxonomy. A  subsequent work, which we will often compare with, is the already cited study by \citet{pitts2014exploratory}, whose research questions stimulated our study.

Our research aims, in particular, we aim to evaluate two research questions:
\begin{itemize}
\item RQ1: the impact of the intervention on the understanding of the role of curvature in explaining gravitational interactions
\item RQ2: the pupils' attitude on learning Einsteinian physics
\end{itemize}
As far as we know, this is the first research study, in an Italian primary school setting, aimed at exploring the basic ideas of Einstenian gravity. The goal of the Einstein First project is to facilitate a widespread implementation of a curriculum which includes the language of modern physics. {As members of this international collaboration, we know that}  results and comparison from different education systems worldwide is fundamental: our study will allow a preliminary evaluation of this effort in the context of Italian primary school.

The paper is organized as follows: in Section \ref{sec:gravity} we discuss some issues in teaching gravity, while in Section  \ref{sec:conme} we focus on the context and the methodologies used in our intervention and the results are in Section \ref{sec:res}.

\section{Issues in teaching gravity}\label{sec:gravity}

In general relativity, space and time are not absolute entities, such as in Newtonian physics, but are part of the four-dimensional continuum, the \textit{spacetime}, which is a ``living'' thing according to  J.A. Wheeler: ``spacetime tells matter how to move, matter tells spacetime how to curve'' \cite{wheeler2000geons}.  Consequently, gravity is explained in terms of a spacetime deformation, due to the presence of matter; in turn, the motion of matter is determined by the spacetime curvature. General relativity includes Newtonian gravity as a particular case: roughly speaking, in the vicinity of each point in the spacetime continuum the deformation is experienced as the force of gravity. According to Newton's law of gravity all bodies having a mass attract each other, and this force is proportional to the  product of the masses, and inversely proportional to the square of their distance; hence, Newton's gravity force is an \textit{action at distance} between bodies. 

Several studies show that students have significant difficulties in understanding gravity. \citet{bar1997children} investigated the ideas about action at distance of students aged between 9 and 17 years: according to some of their conceptions, gravity needs air to act, magnetism and gravity are connected, and the same is true for electrostatics and gravity.  More generally, students need a connection in order to realize the abstract notion of action at distance. 

There is another important issue to be considered when teaching gravity: Newton's  force is symmetrical with respect to the two interacting bodies, since they experience the same and opposite force and this force is universal.  Everyday experience manifestly contradicts this fact, since bodies fall toward the Earth, but the Earth does not move; also, if every body undergoes this kind of attraction, it is difficult to explain why we do not see objects that spontaneously start moving toward each other. It seems that the Earth has a privileged role: \citet{vosniadou1994capturing} showed that primary schools students look at the Earth as a privileged place, with its own laws, and   \citet{baldy2005etude} confirmed that this Aristotelean division between ``the heavens'' and ``the Earth'' remains in 15 yrs old students. According to \citet{posner1982accommodation} a theory, in order to be accepted by students needs to be understandable, plausible, and effective for solving problems: the theory of action at distance does not fullfill these conditions, since the nature of this attraction is not known; moreover it is difficult to take it seriously, since universal attraction between bodies contradict everyday experience. These  remarks show that, as a matter of fact, Newtonian gravity \textit{is not intuitive.}


An exhaustive  reflection on teaching gravity cannot ignore the issue of age appropriate teaching. Taking into account the idea of stages in intellectual development, introduced by \citet{piaget2008psychology}, it is important to consider the concept of \textit{developmentally appropriate practice}, in particular for childhood education. Accordingly, it is important, from a pedagogical perspective, to project and develop teaching activities that meet the students' needs.   
In this context, the use of model and analogies could prove very useful, as suggested by \citet{duit1991role,venville1996role}. These considerations are true, in particular, for teaching gravity; but, probably, explaining Newtonian action at distance is more difficult  than describing gravity as a deformation of spacetime, according to the Einstenian view.

In fact, Einstein's general relativity gets rid of the action at distance problem, since it relates gravitational interaction to the geometry of spacetime. An analogy that can be used to imagine and visualize these geometrical effects is the lycra (or rubber) sheet, where students can see that the attraction between bodies, represented by balls or marbles,  is not the effect of mysterious properties, but just a consequence of the way each body deforms the surface around it. The lycra sheet empty of bodies is the analogue of vacuum spacetime. The presence of a single body deforms the sheet: then, a  nearby object experiences an attractive force as a result of the deformation. This is the analogue of what happens around masses, in spacetime. It is manifest that when two balls are close enough to each other, the surface is deformed by both of them: this emphasizes both the universality of the attraction (each balls deforms the surface around it) and the symmetry in the interaction (the surface is deformed by both balls).

\begin{figure}[ht]
\begin{center}
\includegraphics[scale=1.90]{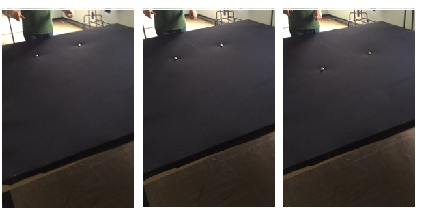}
\includegraphics[scale=1.90]{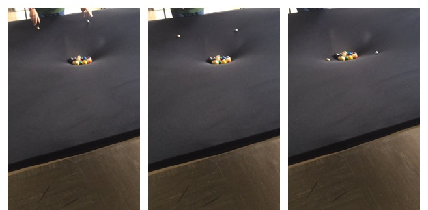}
\caption{The spacetime simulator. Frames on the left show the path of two steel balls on the flat surface of the simulator, which corresponds to propagation in vacuum: the two balls move along straight lines. Frames on the right show the path of the two steel balls on the curved surface of the simulator, around masses: the two balls undergo a deflection with respect to the vacuum case. This is a simple model of gravitational deflection. } \label{fig:simulator}
\end{center}
\end{figure}

This new perspective for teaching gravity, as suggested by \citet{baldy2007new},  {is} based on an analogy which has no claims of physical and mathematical rigour. In fact, the deformation of spacetime is quite different from the deformation of space of the lycra sheet, but it useful for our purposes. The limits of this analogy are well emphasised and discussed by \citet{price2016spatial}: notice that free fall in curved spacetime can be explained as a consequence of the warping of time \cite{gould2016does,stannard2016did2,stannard2018things1,kersting2019free} rather than space. That being said, we believe that this analogy can give a concrete and intuitive visualization of the geometrization of gravity and it can  help students to understand gravitational phenomena and their universality.  In the context of the Einstein First project, the lycra sheet is called the \textit{spacetime simulator}: {the model that we projected and used during our intervention is depicted in Figure \ref{fig:simulator}.}

\section{The Intervention}\label{sec:conme}

The intervention was carried out in three ending classes of primary schools in the city area of Turin, and it lasted 15 hours. The classes are located in different socio-cultural contexts. Two classes, composed of 21 and 22 pupils, belongs to ``Tetti Francesi School'', located in an industrial area in the Turin suburbs, with a strong presence of the working-class and a little percentage of foreign pupils and gipsy ethnic groups. The third class, composed of 20 pupils,  belongs to the ``Buon Consiglio School'', which is a catholic institution  located in a wealthy neighbourhood in the city centre.  Consequently, the class level was medium-high.   {Overall, 58 students (28 males and 30 females)  aged between 10 to 11 years were involved in the whole trial,  {which was entirely undertaken by one of us (S. M.), in collaboration with the schools' teachers, whose role was} to facilitate and stimulate the participation of the students to the activities proposed.}  

As we said before, the whole research study was aimed at introducing some fundamental aspects of the theory of relativity to pupils who were uneducated in the Newtonian physics; in particular, here we focus on  the Einsteinian idea of gravity as a geometric property of  spacetime, deformed by the presence of masses. To emphasize this aspect we focused on the concept of curvature. In primary school it is impossible teaching modern physics using maths formulas: hence,   the whole intervention was developed through the use of language and, also, we exploit the use of model and analogies. 
 {Finally,} in our intervention we focused also on the historical context in which Galileo and Einstein lived and made their fundamental discoveries.

Our intervention started with a questionnaire administered two weeks before the beginning of our activities;  one month after the end of the activities, we administered a second questionnaire with the goal of studying the delayed impact of our actions.

 The pre-intervention questionnaire contains many Einstein First project's items (e.g. \cite{kaur2017teaching3}). In particular, these are the relevant questions:
\begin{itemize}
\item According to you, what is gravity?
\item Can the sum of the angles in a triangle be different from 180 degrees?
\item Can parallel lines ever meet?
\end{itemize}
The final questionnaire is different, since the items are more focused on the specific topics covered during the intervention, which were  tailored to meet particular pedagogical and didactic necessities. It is possible, however, to compare the pre and post answers after having identified the main topic. So, the relevant questions present in the final questionnaire are
\begin{itemize}
\item What happens to the space when there are very heavy masses such as the Sun? Remember the experiment with the lycra sheet.
\item What happens to parallel lines and to the inner angles of a triangle on curved surfaces? Why?
\end{itemize}

In the first part of the laboratory activities, we used the spacetime simulator;   the activity develops as follows: first, the teacher invites pupils to sit around the structure and to observe it, then she introduces some challenging questions. In this case, we noticed that some  basic astronomy notions could be very useful for the pupils   to better understand the phenomena discussed.  Observation and meta-reflections are essential in this phase: the teacher helps the pupils to formulate hypotheses, to describe the situation and, finally, to understand, with the help of the marble motions on the lycra sheet, the interplay between masses in the Solar System. To begin with, pupils observe the motion of a single marble on the flat empty sheet, and deduce that the trajectory is a straight line; then, two marble moves in straight lines. The same motions are subsequently observed on the sheet where huge masses  were put to curve it, and the pupils observe that the trajectories are not any longer straight:  marbles move around huge masses just like planets around the Sun; the greater the masses, the more curved the marbles trajectories (see figure \ref{fig:simulator}).  The latter fact enables the teacher to introduce  the discussion about black holes, which are presented as extremely massive objects, and pupils are asked to make their hypotheses about motion around these objects, and the meaning of time in the vicinity of them.   Usually pupils are very attracted by the mysteries of the Universe, even though they cannot really understand the theoretical principles. 

\begin{figure}[h]
\begin{center}
\includegraphics[scale=.50]{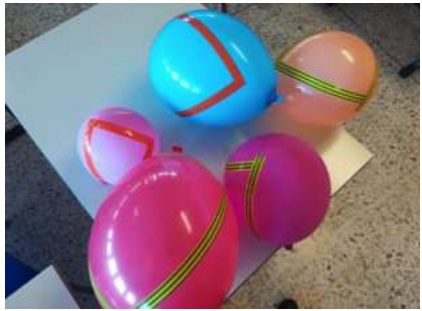}
\caption{Triangles on the surface of balloons: pupils are asked to measure their angles.} \label{fig:palloncini}
\end{center}
\end{figure}

In the second part of the laboratory activities, we focused on the properties of curved surfaces and, in particular, on the sum of the inner angles of triangles. In this case, pupils are asked to work in pair:  each pair receives a balloon, of different sizes, with a triangle drawn on it, and they are asked to measure the interior angles (see Figure \ref{fig:palloncini}). These measurements are compared to what happens on a flat surface. A discussion arises on the analogy with what happens on the curved surface of the spacetime simulator: in particular, the facts that parallel lines can meet and the sum of interior angles can be greater than 180 degrees  are focused on.

%

\section{Results and Discussion}\label{sec:res}

\begin{figure}[h]
\centering
\includegraphics[scale=.40]{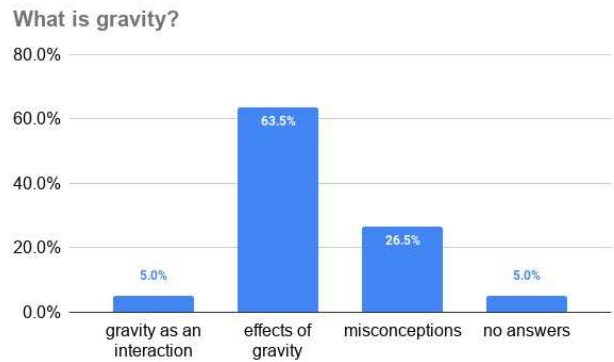}
\caption{Answers to the question:  \textit{according to you, what is gravity?}}\label{fig:1r}
\end{figure}

The  pre-intervention questionnaire addressed the meaning of gravity (\textit{according to you, what is gravity?}). The answers were somewhat heterogenous, hence we divided them in categories referring to different awareness levels (see Figure \ref{fig:1r}):
\begin{itemize}
\item {gravity as interaction} (5\%): the students explain gravity as a force of interaction between masses, instead of a force which acts unidirectionally; for instance: \textit{``The gravity force depends on mass of the bodies, thus for me it means the attraction between two bodies.''} 
\item {effects of gravity} (63.5\%): many students  describe gravity in connection with its effects on everyday life, as they can learn from books: \textit{``Gravity is when we stay attached on the ground''.}; \textit{``Gravity is like a magnet which allows us to remain attached to the ground''}; \textit{``Gravity force makes us stay fixed on the ground, otherwise we would fly in the space''}
\item {misconceptions} (26.5\%): we collect in this group answers with conceptual mistakes, such as \textit{``For me  gravity force is when we fly in the air''}
\item {no answer} (5\%): few students  do not answer
\end{itemize}

 {It is interesting to emphasise that a few students describe gravity as interaction,}  thus evidencing that they are aware of the Newtonian definition of gravitational force. Unsurprisingly, no reference is made to the modern view of gravity as a geometric property of space-time.

\begin{figure}[h]
\centering
\includegraphics[scale=.35]{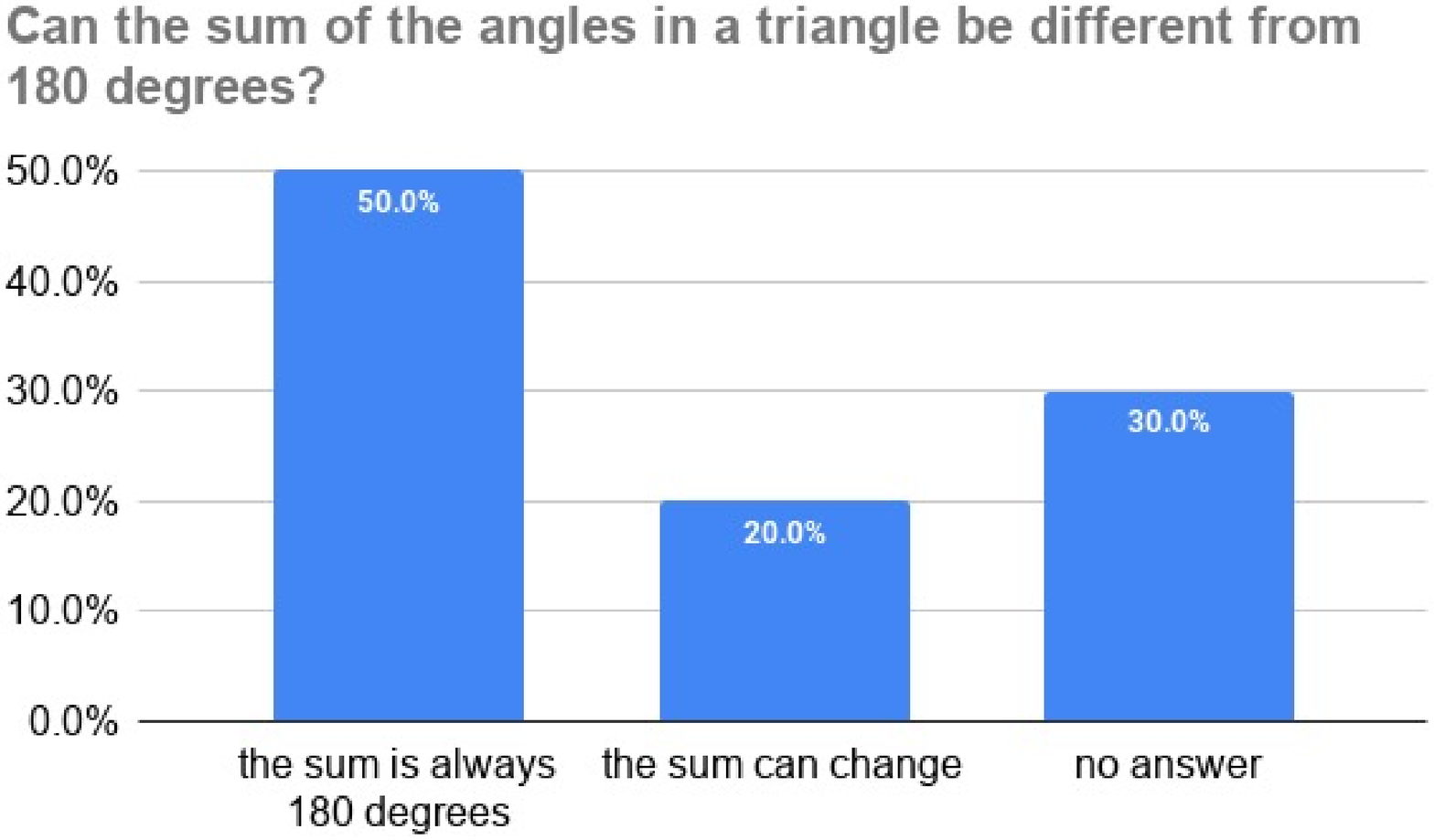}
\includegraphics[scale=.35]{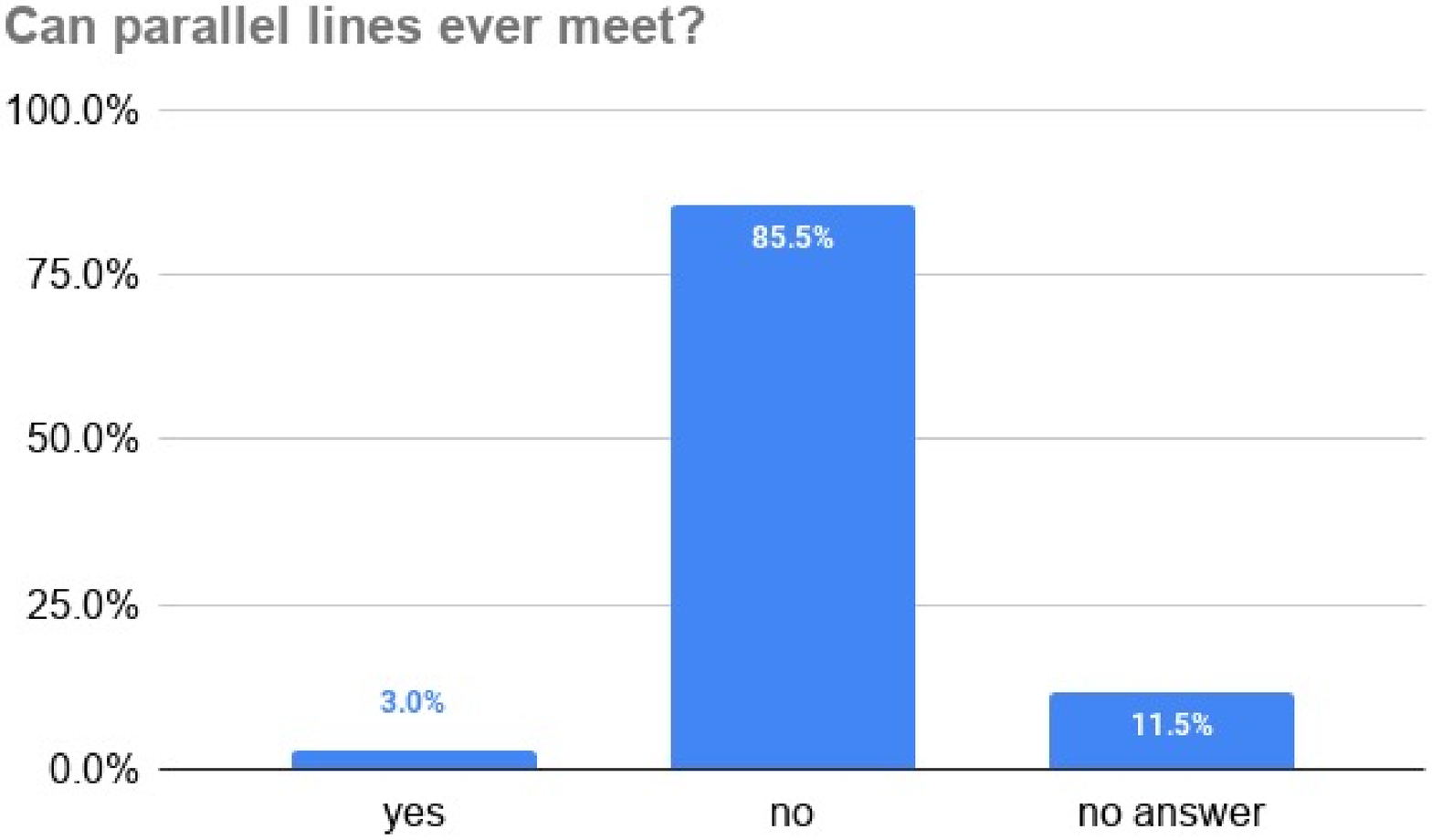}
\caption{Left: answers to the question  \textit{Can the sum of the angles in a triangle be different from 180 degrees?}; Right:  answers to the question \textit{Can parallel lines ever meet?}}\label{fig:23r}
\end{figure}

Since gravity is the geometry of spacetime in general relativity, we tested the knowledge of basic facts of Euclidean geometry. We started from the known results pertaining to the sum of internal angles of a triangle, which, according to Euclidean geometry, needs to be 180 degrees. 
{Unfortunately, even though this topic is included in the Italian curriculum of primary school  it was studied by only one class in our sample}. The results are in Figure \ref{fig:23r}. If, on the one hand, we see that the majority of students remembers this rule, it unlikely that the wrong answers are determined by the knowledge of non Euclidean geometries: perhaps  those students simply do not remember or know the rule. 
Things are quite different for another rule of Euclidean geometry, that is the fact that two parallel lines can never meet. Indeed, this result is very well known also from everyday life (\textit{``Parallel lines cannot meet because they are as railways''}) even though students do not know the formulation of Euclid's fifth postulate. The results for the answer \textit{Can parallel lines ever meet?} are in Figure \ref{fig:23r}.
The results of our pre-intervention questionnaires may be summarised as follows: our pupils have a poor understanding of gravity, at least as an interaction between bodies; rather, they are quite familiar with the effects of gravity; as for the basic facts of Euclidean geometry, our pupils show to have a good level of understanding. The subsequent activities  have the role to provoke a conflict between these basic knowledges and the view of modern physics.

\begin{figure}[h]
\centering
\includegraphics[scale=.35]{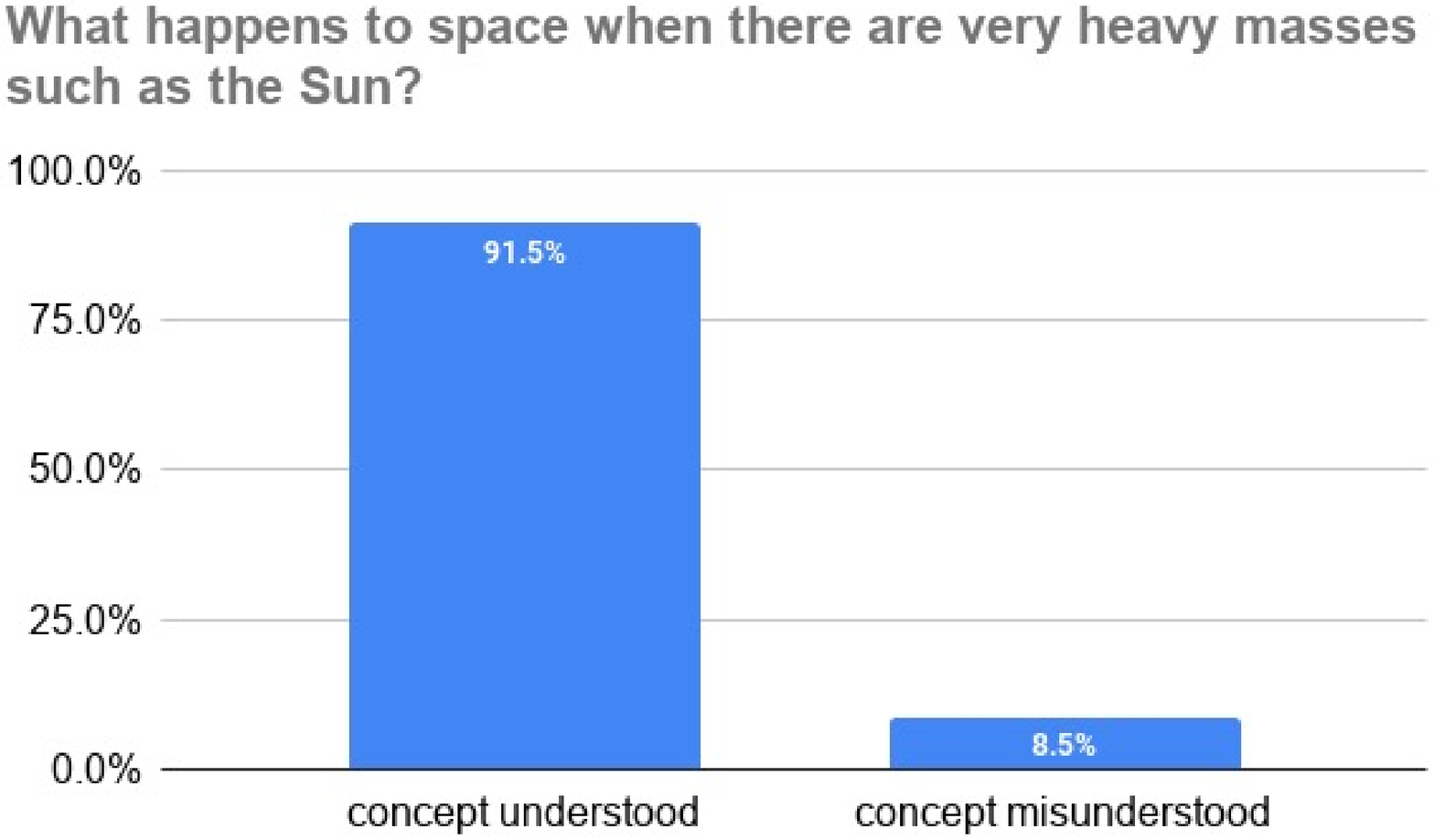}
\includegraphics[scale=.35]{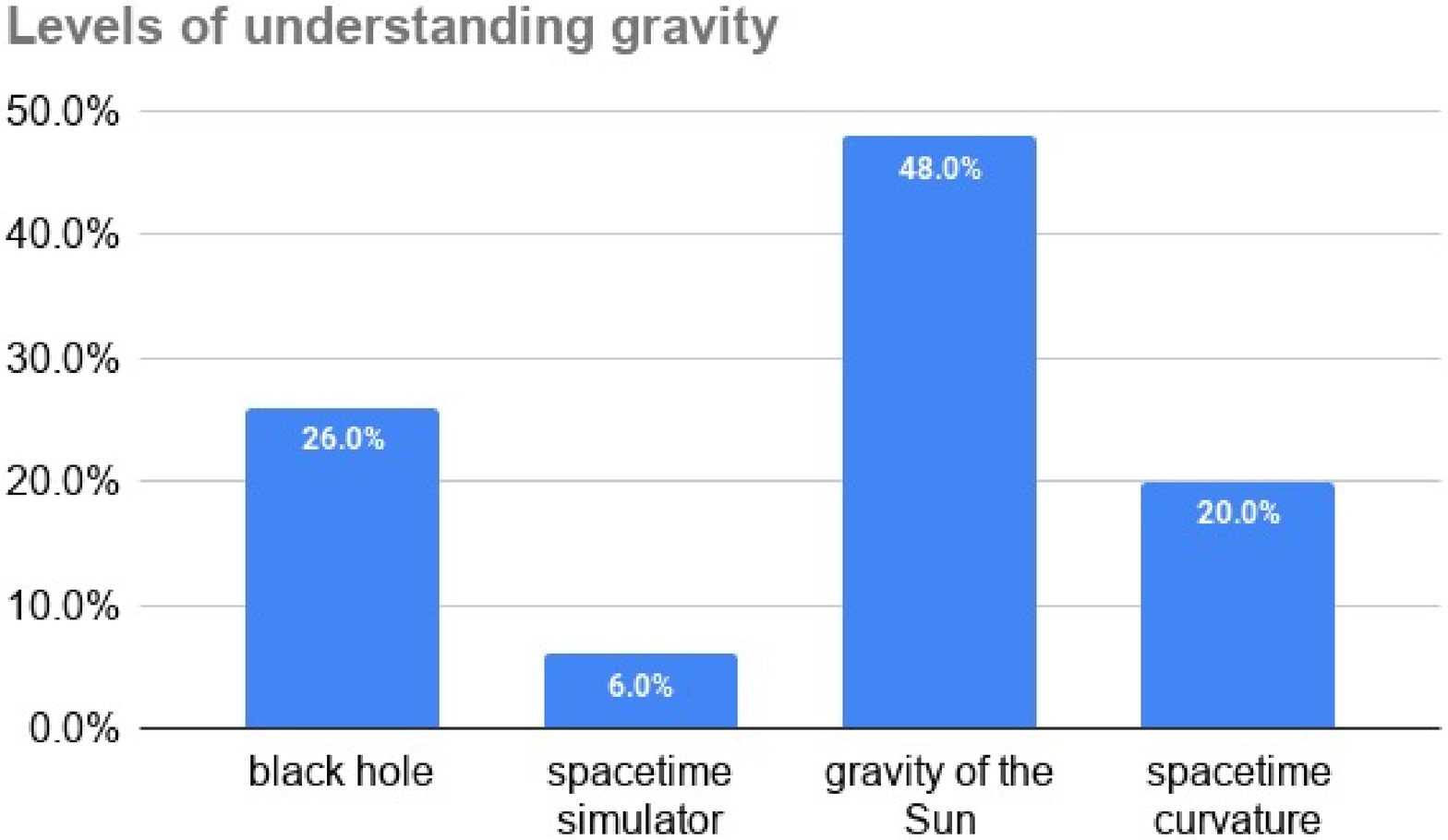}
\caption{Left: answers to the question \textit{What happens to space when there are very heavy masses such as the Sun? Remember the experiment with the spacetime simulator.} Right: different levels of understanding gravity } \label{fig:45r}
\end{figure}

After the diagnosis of prior knowledge phase, we used the spacetime simulator to explain the Einsteinian view of gravity.  In particular, we showed that huge mass can deform the lycra sheet and, thus, mass moving in this deformed space are forced to follow curved trajectories and we emphasized the analogy with the motion of the planets around the Sun.  The answers to the question \textit{what happens to space when there are very heavy masses such as the Sun? Remember the experiment with the spacetime simulator} are in Figure \ref{fig:45r}. The results show that  almost all pupils understand the experiment with the spacetime simulator and its analogy with celestial bodies. 

However, a deeper analysis emphasizes different levels of understanding. In particular (see Figure \ref{fig:45r}):
  \begin{itemize}
\item 26\% of the answers explicitly relate spacetime curvature to black holes: \textit{``Since it is an heavy mass, it creates a black hole''; ``They  gather to as long as they make a black hole''; ``The Sun attracts other planets, the heavy weight of  all masses drill the spatial spacetimes simulator and create a black hole''.} These answers clearly show that students are fascinated by black-holes and their mysteries. 
\item  6\% of the students limit themselves to describe the experiment, using the proper terminology of celestial bodies: \textit{``We put one billiard marble in the centre of the spacetime simulator and it was the Sun and then   other lighter marbles roll over the simulator and the Sun attract them (the lighter marbles were the planets)''}
\item 48\% of the students, that is to say the great majority, focus on the role of Sun and mention the attraction due to its mass: \textit{``Lighter masses  are attracted by the heavier one''; ``Masses accumulate and attract other planets which move around them''; ``The other celestial bodies cannot move on straight lines, but they move around the Sun because its mass  attracts them''}
\item  20\% of the students' answer explicitly mention the spacetime curvature: \textit{``The Sun curves the space, thus it attracts other corps as the Earth''; ``If there are heavy masses as the Sun, space is curved''; ``It happens that black holes will be created because of space deformation''}
\end{itemize}
We see that more than 50\% of the students give explanations in terms of modern physics: both the concepts of spacetime curvature and black holes, in fact, refers to the Einsteinian view. More in general, the analogy seems to be useful to explain the interaction between bodies, without difficulties deriving from the need to explain action at distance. 

\begin{figure}[htp]
\centering
\includegraphics[scale=.45]{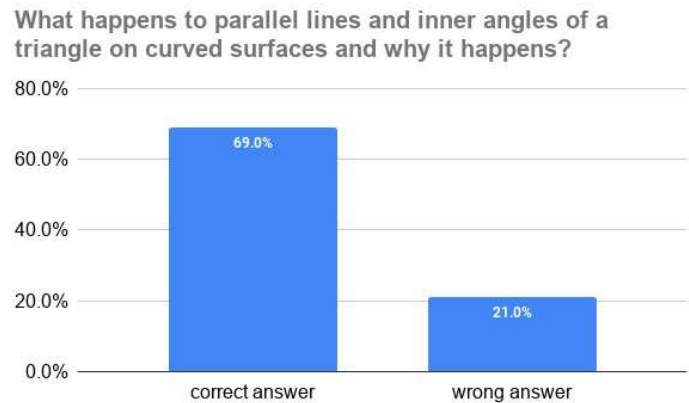}
\caption{Answer to the question: \textit{What happens to parallel lines and inner angles of a triangle on curved surfaces and why it happens?} } \label{fig:6r}
\end{figure}

After the intervention, students seem to have correctly understood the ``strange'' properties of curved surfaces, as it is shown in the results for   the question \textit{what happens to parallel lines and inner angles of a triangle on curved surfaces and why it happens?}, summarised in Figure \ref{fig:6r}. In particular, the experiments at the spacetime simulator enable to relate the presence of masses to deformations which, in turn, influence the motion of other masses, which do not follow straight lines. In particular, we see that  70\% of the answers are correct and correctly motivated: \textit{``They meet because the Universe is not flat; the sum of inner angles is not 180 degrees because the surface is modified''; ``The angles are modified because of the curvature of the surface; in fact, they are not the same. So geometry in space is different from the one on Earth'';  ``Parallel lines in space are no more parallel because they meet, while the sum of internal angles on a curved space is not 180 degrees:   this rule is no more valid''.}
Even though comparison with the results of pre-intervention questionnaire (see Figures \ref{fig:23r}) show a marked improvement,  as expected, students  have difficulties in the process of abstraction. In fact,  they  show to have understood the correlation between mass and motion but just a few of them manage to perceive explicitly gravity as a property of spacetime. This behaviour could be explained through Piaget's developmental levels \cite{piaget2003psychology}. In particular, among the correct answers, the great majority (70\%) are correct but without interpretation, while in the rest (30\%) students give a suitable scientific  interpretation (e.g. \textit{``Since it has a heavy mass, the Sun deforms the space and makes the gravity''.}) 

\begin{figure}[htp]
\centering
\includegraphics[scale=.30]{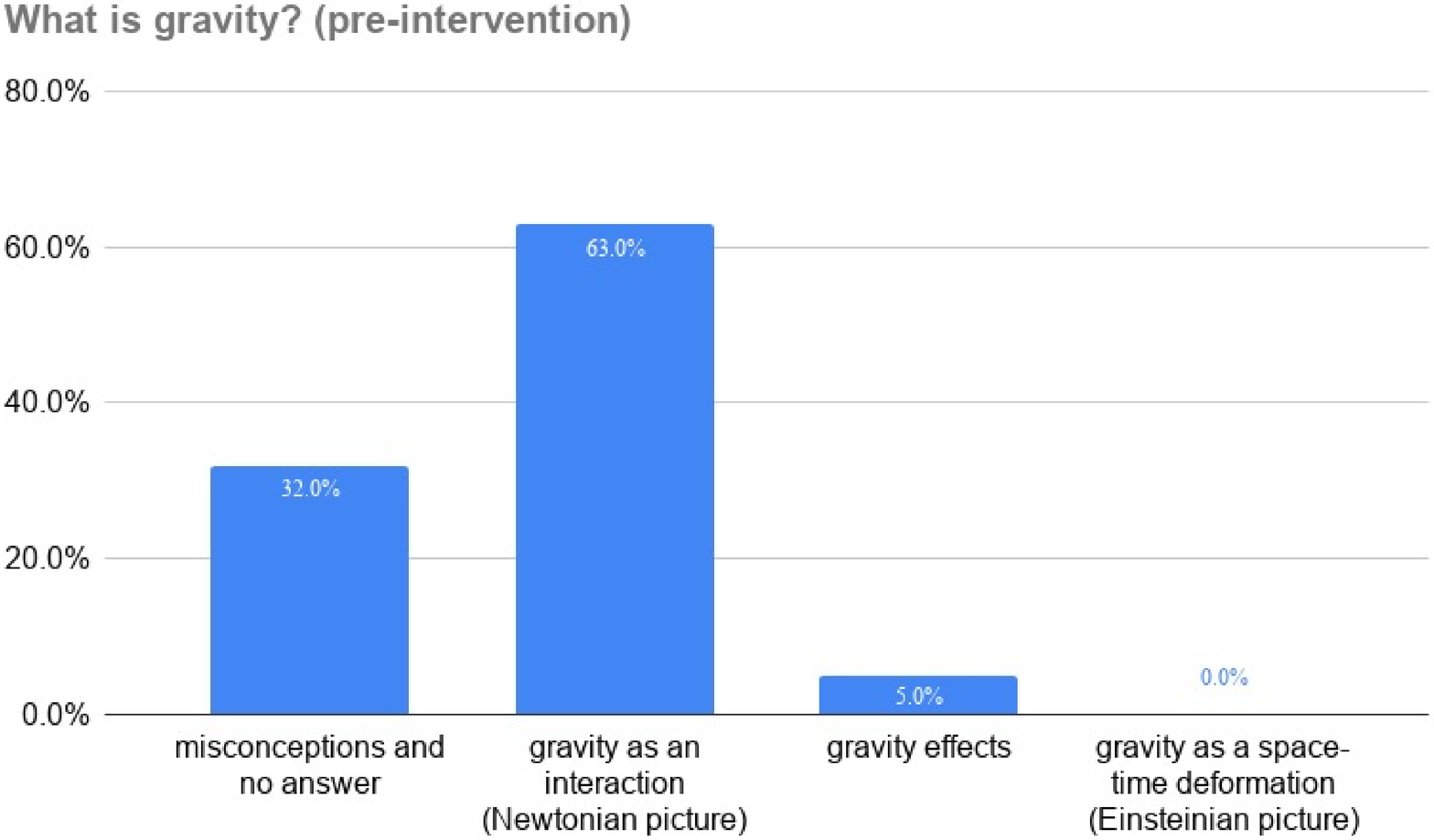}
\includegraphics[scale=.30]{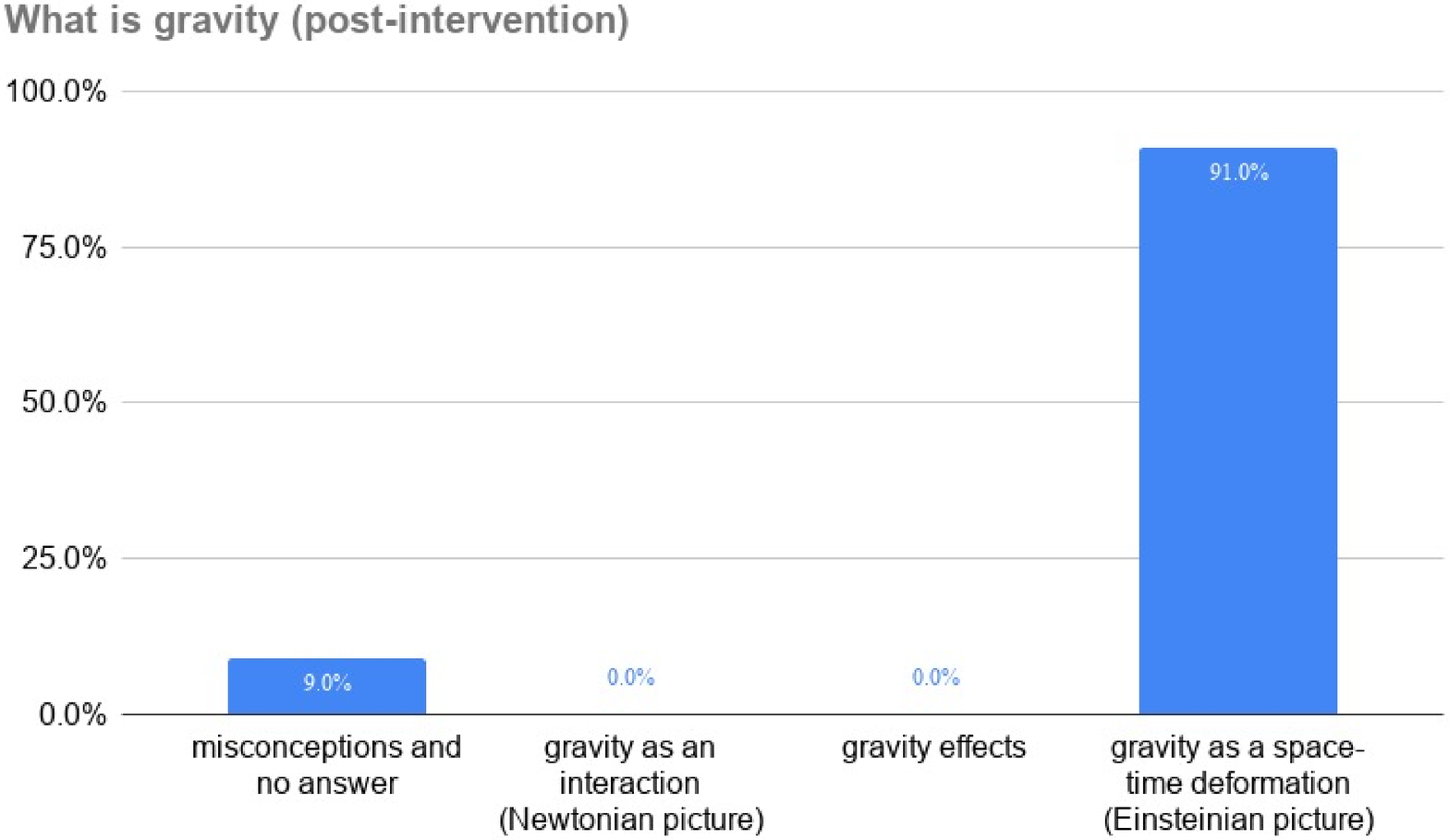}
\caption{Old and new ideas about gravity. } \label{fig:78r}
\end{figure}

In Figure \ref{fig:78r} we draw a comparison between the meaning of gravity, as perceived by the pupils before and after the intervention, using a different categorization  to emphasize the perception of the Newtonian and Einsteinian approaches. 
As it is possible to see from the histograms, what pupils think about gravity is considerably changed: in fact, contrary to the pre-intervention answers, where 63\% described gravity in terms of effects on everyday life, in the post-intervention questionnaire  students refer to gravity as an interaction or a force between bodies.  Although the concept of deformation is often implicit, many students mention the attraction due to \textit{weight} of the bodies. {These are some answers to the question (which refers to the space-time simulator experiment) \textit{Why the masses meet?}: \textit{Because of their weight}; \textit{each mass is attracted by the  weight of the other, they meet and create a basin}; \textit{since it has a heavy mass, the Sun deforms the space and makes the gravity}} In this case, it is  unlikely that they refer to a strictly Newtonian definition, rather it could be possible that, since they may have difficulties in understanding gravity as geometric property according the the Einsteinian picture, they correctly understand the meaning of attraction because of what they learned at the spacetime simulator, where \textit{heavy bodies} deform the sheet. 

{It is important to stress that we chose to report  aggregated results for the three classes involved, since there are no significant differences, notwithstanding  the sharp distinction  between the socio-cultural contexts of the two schools.}

\begin{figure}[htp]
\centering
\includegraphics[scale=.1]{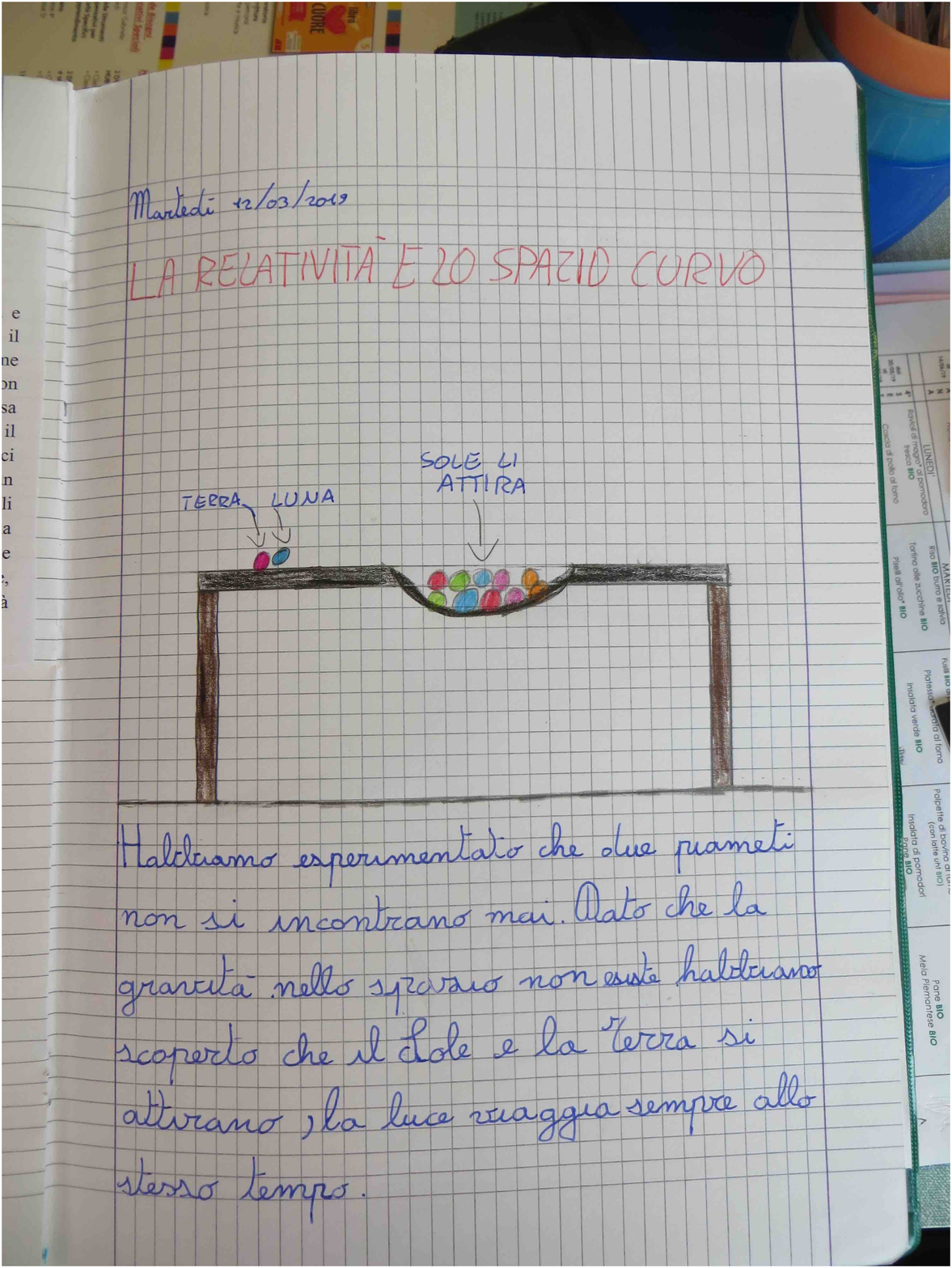}
\includegraphics[scale=.1]{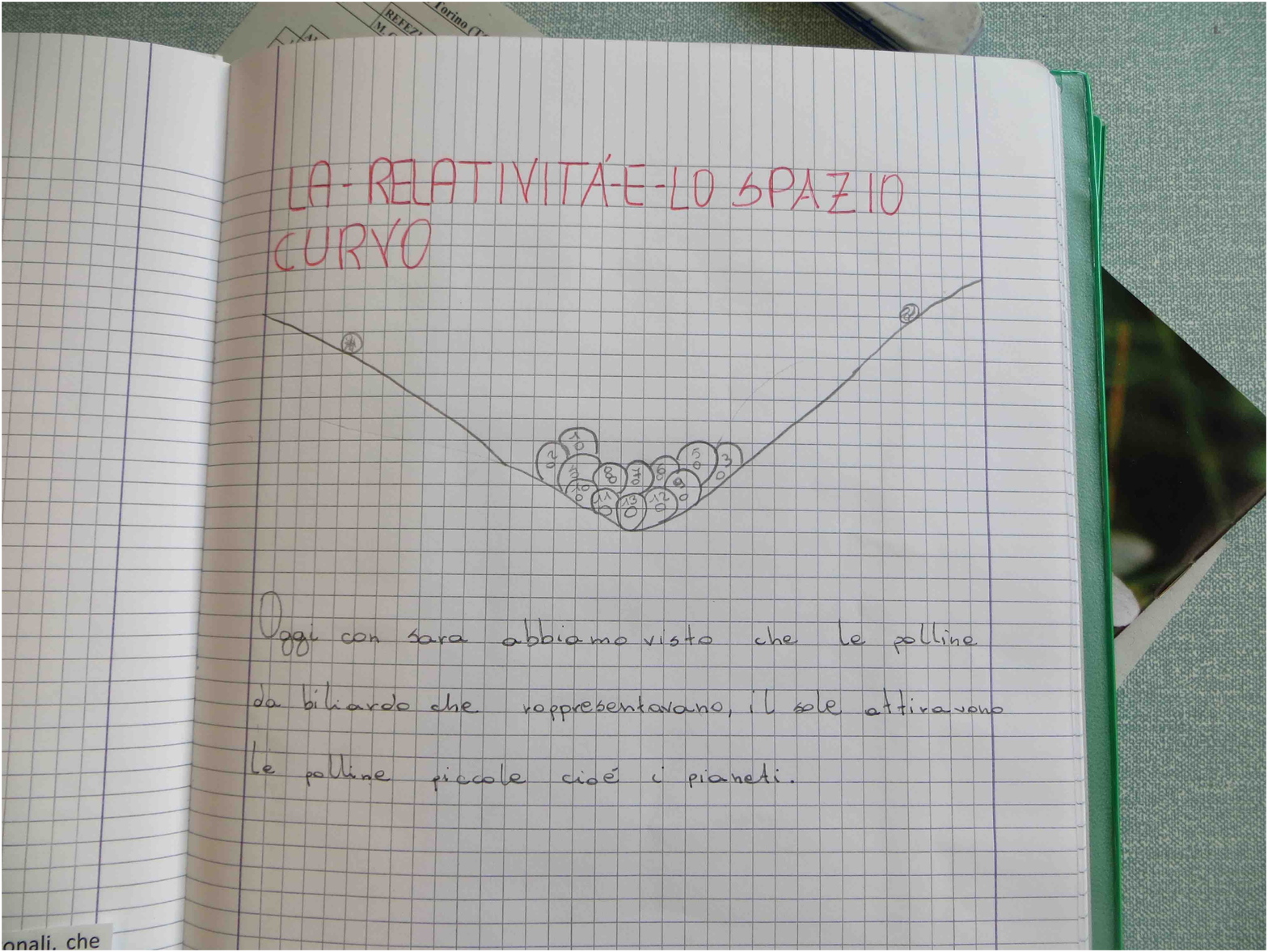}
\caption{From the pupils' notebooks: the deformation of the spacetime simulator.} \label{fig:S12}
\end{figure}

In conclusion, with regard to the first research question (RQ1) our results show that while before our intervention the great majority of students explained gravity in terms of its effects in everyday life, after our intervention they mostly refer to  gravity as a deformation, according to the Einsteinian picture; these results are relevant also in terms of retention of the key ideas, since the second questionnaire was administrated one month after the end of the activities.  It is interesting to point out that the great majority of students  referred to gravity in terms of interaction between bodies: in this case, is likely to think that the concept of deformation that they learned at the spacetime simulator is implicit in their idea of interaction.  The perception of gravity as deformation is beautifully depicted in the pupils' notebooks, as we may see in Figure \ref{fig:S12}.

Our results suggest not only that is possible to teach Einsteinian physics in primary school but, also, that it proves useful: in fact,  thanks to the possibility of visualizing them, the basic ideas of general relativity can be naturally thought using analogies. Students enjoyed a lot the activities proposed (100 of positive answers in the questionnaire) which helped to drive their interest in these new topic: this is a very good answer to our second research question (RQ2). 

{Our work can be thought of as a replication, in a different context, of previous results obtained in Australian schools, and it contributes to consolidate the effectiveness of this approach. Seemingly, there are no effects deriving from the diverse socio-cultural contexts where the intervention were performed and from the different education systems and teachers' attitudes.}

 {Following common sense, we usually assume modern physics} to be more counter-intuitive and hard for a child than it really is, without considering that adults already have a precise mindset that children don't have. These are just preliminary results and a first step toward the possibility of introducing modern physics in schools: further investigations are needed also to design a vertical curriculum.  However, these results show also that it is not impossible to include in the traditional education programs the basic ideas of modern physics and to integrate them with the classical physics paradigm, to give students a comprehensive view of physics starting from the early age. This approach could allow  to give all people, and not only to an elite, the possibility of understanding these beautiful intellectual achievements.

\noindent \textbf{Ethical Statement:} Data used in this research were collected during an undergraduate thesis project. Pupils involved were informed about the treatment of data for research purposes.

\end{document}